\documentclass[prd,showpacs,twocolumn,nofootinbib,superscriptaddress,preprintnumbers,amsmath,amssymb]{revtex4-1}
\usepackage{tikz-feynman}

\usepackage{url}
\usepackage{amsfonts,multirow}
\usepackage{rotating}
\usepackage{color}
\usepackage{hhline}
\usepackage{slashed}

\usepackage{xcolor}

\usepackage{array,multirow}

\usepackage{framed}
\usepackage{bm}
\usepackage{amsmath}
\usepackage{graphicx}
\usepackage{hyperref}
\usepackage{subfigure}
\usepackage{epstopdf}
\usepackage{verbatim} 
\usepackage{array}
\usepackage{booktabs}
\usepackage{color}

\pdfoutput=1

\definecolor{Gray}{gray}{0.85}
\definecolor{LightGreen}{rgb}{0.88, 1, 0.88}
\definecolor{Lime}{rgb}{0,255,0}
\definecolor{LightCyan}{rgb}{0.88,1,1}
\definecolor{LightRed}{rgb}{1, 0.85, 0.85}
\definecolor{Red}{rgb}{1, 0, 0}
\definecolor{LightYellow}{rgb}{1, 1, 0.85}
\definecolor{Yellow}{rgb}{1,1,0.05}
\definecolor{LightBlue}{rgb}{0.87, 0.94, 1}
\definecolor{white}{gray}{1}
\definecolor{black}{gray}{0}
\newcolumntype{G}{>{\columncolor{LightGray}}c}

\def\beq{\begin{equation}}
\def\eeq{\end{equation}}

\def\bea{\arraycolsep .1em \begin{eqnarray}}
\def\eea{\end{eqnarray}}

\def\eq#1{(\ref{#1})}

\def\s0#1#2{\mbox{\small{$ \frac{#1}{#2} $}}}
\def\0#1#2{\frac{#1}{#2}}

\def\grgl{\:\hbox to -0.2pt{\lower2.5pt\hbox{$\sim$}\hss}{\raise3pt\hbox{$>$}}\:}
\def\klgl{\:\hbox to -0.2pt{\lower2.5pt\hbox{$\sim$}\hss}{\raise3pt\hbox{$<$}}\:}

\newcommand \be {\begin{equation}}
\newcommand \ee {\end{equation}}
\newcommand \bed {\begin{displaymath}}
\newcommand \eed {\end{displaymath}}

\newcommand{\bit}{\begin{itemize}}
\newcommand{\eit}{\end{itemize}}

\usepackage{colortbl}

\definecolor{Gray}{gray}{0.85}
\definecolor{LightGray}{gray}{0.93}
\definecolor{LightGreen}{rgb}{0.88, 1, 0.88}
\definecolor{LightCyan}{rgb}{0.88,1,1}
\definecolor{LightRed}{rgb}{1, 0.85, 0.85}
\definecolor{LightRed}{rgb}{1, 0.88, 0.88}
\definecolor{LightYellow}{rgb}{1, 1, 0.85}
\definecolor{LightBlue}{rgb}{0.87, 0.94, 1}
\definecolor{white}{gray}{1}

\usepackage{array,mathtools,amssymb,booktabs}
\newcolumntype{C}{>{$}c<{$}}
\AtBeginDocument{
\heavyrulewidth=.16em
\lightrulewidth=.1em
\cmidrulewidth=.03em
\belowrulesep=.4ex
\belowbottomsep=0pt
\aboverulesep=.4ex
\abovetopsep=0pt
\cmidrulesep=\doublerulesep
\cmidrulekern=.5em
\defaultaddspace=.5em
}

\makeatletter
    \def\CT@@do@color{%
      \global\let\CT@do@color\relax
            \@tempdima\wd\z@
            \advance\@tempdima\@tempdimb
            \advance\@tempdima\@tempdimc
    \advance\@tempdimb\tabcolsep
    \advance\@tempdimc\tabcolsep
    \advance\@tempdima2\tabcolsep
            \kern-\@tempdimb
            \leaders\vrule
                    \hskip\@tempdima\@plus  1fill
            \kern-\@tempdimc
            \hskip-\wd\z@ \@plus -1fill }
            
\begin{document}

\preprint{DO-TH 19/15}
\preprint{QFET-2019-09}

\title{Anomalous Magnetic Moments from Asymptotic Safety}

\author{Gudrun Hiller}
\author{Clara~Hormigos-Feliu}
\affiliation{Fakult\"at Physik, TU Dortmund, Otto-Hahn-Str.4, D-44221 Dortmund, Germany}
\author{Daniel~F.~Litim}
\author{Tom~Steudtner}
\affiliation{Department of Physics and Astronomy, University of Sussex, Brighton, BN1 9QH, U.K.}

\begin{abstract}
The measurements of the muon and electron anomalous magnetic moments  hint at 
physics beyond the standard model. 
We show why   and how 
models inspired by asymptotic safety can explain deviations from standard model predictions  naturally. Our setup features an enlarged scalar sector and  Yukawa couplings between   leptons and new vector-like fermions. Using the complete two-loop running of couplings, we  observe a well-behaved high energy limit of models including  a stabilization of the  Higgs. We find that a manifest breaking of lepton universality beyond standard model Yukawas  is not necessary to   explain the muon and electron anomalies.
 We  further predict the tau anomalous magnetic moment, and  new particles in the TeV energy range whose signatures at
colliders are indicated.  With small CP phases, the electron EDM can be as large as the present bound.
\end{abstract}

\maketitle

{\it Introduction.---} Measurements of the electron and muon anomalous magnetic moments  exhibit intriguing discrepancies from   standard model (SM) predictions
\cite{Tanabashi:2018oca, Hanneke:2008tm,Parker:2018vye}. Adding uncertainties in quadrature, the deviations
\begin{equation}\label{eq:deltagminus2}
\begin{array}{rl}
\Delta a_{\mu} & \equiv a_{\mu}^{\text{exp}} - a_{\mu}^{\text{SM}} = \ 268(63)(43)\cdot 10^{-11}\,, \\[1ex]
\Delta a_{e}& \equiv a_{e}^{\text{exp}} - a_{e}^{\text{SM}} = -88(28)(23)\cdot 10^{-14}
\end{array}
\end{equation}
 amount to $3.5 \, \sigma$ ($2.4 \, \sigma$) for the muon (electron). Recent theory predictions for $a_\mu$ find up to $4.1 \, \sigma$ \cite{Jegerlehner:2017lbd,Davier:2016iru}. 
There are two stunning features in the data. First, the deviations $\Delta a_{\mu} $ and $\Delta a_{e} $ have opposite sign. Second,  their ratio $\Delta a_{e}/\Delta a_{\mu} =-(3.3 \pm 1.6) \cdot 10^{-4}$  is an order of magnitude smaller than the lepton mass ratio $m_e/m_\mu$ and an order of magnitude larger than  the square of the mass ratio $(m_e/m_\mu)^2$. 
Theory explanations of the data \eqref{eq:deltagminus2} 
with either new light scalars \cite{Davoudiasl:2018fbb,Liu:2018xkx,Gardner:2019mcl,Bauer:2019gfk}, supersymmetry \cite{Dutta:2018fge,Endo:2019bcj,Badziak:2019gaf}, bottom-up models \cite{Crivellin:2018qmi,Crivellin:2019mvj}, or other \cite{Han:2018znu,Abdullah:2019ofw},
manifestly break  lepton flavor universality.

In recent years, asymptotic safety 
has  been put forward as a new idea for model building
 \cite{Bond:2017wut,Kowalska:2017fzw}. It is based on the discovery \cite{Litim:2014uca} that particle theories may very well remain  fundamental and predictive  in the absence of asymptotic freedom due
 to 
  interacting
  high energy fixed points 
  \cite{Wilson:1971bg,Bailin:1974bq,Weinberg:1980gg}. 
For  weakly coupled theories, general theorems for  asymptotic safety are available \cite{Bond:2016dvk,Bond:2018oco} with templates covering simple \cite{Litim:2014uca,Bond:2017tbw,Bond:2019npq}, 
semi-simple \cite{Bond:2017lnq}, and supersymmetric gauge theories \cite{Bond:2017suy}. 
Yukawa interactions and new scalar fields play a prominent role because 
they  slow down the growth of asymptotically non-free gauge couplings, which can enable interacting fixed points \cite{Bond:2016dvk} including in extensions of the standard model
\cite{Bond:2017wut,Kowalska:2017fzw,Barducci:2018ysr,Hiller:2019tvg}.

In this Letter, 
we show that
 asymptotically safe extensions of the SM
may offer  a natural  explanation for the data
 \eq{eq:deltagminus2}.
The primary reason for this is that   Yukawa interactions, which  help generate interacting fixed points, 
  can {\it also} contribute to lepton anomalous magnetic moments. 
We demonstrate this idea  in 
two concrete models  
by introducing  Yukawa couplings between ordinary leptons and  new vector-like fermions, and by adding new scalar fields which admit either a flavorful or flavor universal ground state. Unlike in all previous works \cite{Davoudiasl:2018fbb,Liu:2018xkx,Gardner:2019mcl,Bauer:2019gfk,Dutta:2018fge,Endo:2019bcj,Badziak:2019gaf,Crivellin:2018qmi,Crivellin:2019mvj,Han:2018znu,Abdullah:2019ofw}, we  find that the data \eq{eq:deltagminus2} can be accommodated  without any explicit breaking of lepton-universality.
The stability of SM extensions 
all the way up to the Planck scale is exemplified using  the renormalization  
group (RG) running of couplings for a wide range of BSM parameters.

{\it New vector-like fermions and scalar matter.---}
In the spirit of  \cite{Litim:2014uca}, we are interested in SM extensions  involving $N_F$ flavors of vector-like color-singlet fermions $\psi_i$ and $N_F^2$ complex scalar singlets $S_{ij}$.  In their simplest form, the new fermions couple to  SM matter only via  gauge interactions  \cite{Bond:2017wut,Kowalska:2017fzw}. The   new ingredient in this letter  are   Yukawa couplings between SM and BSM matter. To make contact with SM flavor we set $N_F=3$. We then consider  singlet or doublet models where the new fermions  are  either $SU(2)$ singlets with hypercharge $Y=-1$,  or    $SU(2)$ doublets with $Y=-\frac12$. In our conventions, electric charge $Q$ and weak isospin  $T_3$ relate as $Q = T_3 + Y$. Within these choices, and denoting the  SM lepton  singlets, doublets and Higgs as $E, L$ and $H$, respectively,  we find three  possible Yukawa couplings $\kappa$, $\kappa'$ and $y$ with 
 \begin{equation} \label{eq:Yukawa}
\begin{array}{l}
\!\!\mathcal{L}^{\text{singlet}}_{\text{Y}} = -\kappa \overline{L} H \psi_R  - \kappa'\overline{E}S^{\dagger}\psi_L - y\, \overline{\psi}_L S \psi_R + \mathrm{h.c.}\\[1ex]
\!\!\!\!\mathcal{L}^{\text{doublet}}_{\text{Y}} = -\kappa \overline{E} {H}^{\dagger} \psi_L - \kappa'\, \overline{L}S\psi_R - y\, \overline{\psi}_L S \psi_R + \mathrm{h.c.}
\end{array}
\end{equation}
and flavor traces are understood to simplify the subsequent RG analysis.
Effects of the Yukawa coupling $y$ have been studied in \cite{Bond:2017wut,Kowalska:2017fzw,Barducci:2018ysr}.
The scalar potential  of either model reads
\begin{equation}\label{eq:scalarV}
\begin{aligned}
V = &  \ \ \ \lambda\, (H^{\dagger}H)^2 + \delta \, H^{\dagger}H\,{\rm Tr}\left[S^{\dagger}S\right] \\
&+ u\, {\rm Tr}\left[S^{\dagger}SS^{\dagger}S\right]
+ v \, \left({\rm Tr} \left[S^{\dagger}S\right]\right)^2 \,,
\end{aligned}
\end{equation} 
where $u,v,\lambda$ and $\delta$ are  quartic and portal couplings. We further introduce  mass terms for the scalars and vector-like fermions. 
The potential  (\ref{eq:scalarV}) admits vacuum configurations  $V^+$ and $V^-$ 
characterized by
\begin{equation}\label{eq:vstab}
\begin{aligned}
V^+ : \quad & 
\left\{ \begin{array}{l}
\lambda > 0,\quad u > 0,\quad  u + 3\,v > 0,  \\
\delta >  -2 \sqrt{\lambda\left(u/3 + v\right)}\,,
\end{array} \right. \\
V^- :  \quad &
\left\{ \begin{array}{l}
\lambda > 0,\quad u < 0,\quad  u + v > 0, \\
\delta >  -2 \sqrt{\lambda\left(u + v\right)}\,.
\end{array} \right. \\
\end{aligned}
\end{equation}
Either of these  allows for  electroweak symmetry breaking. 
Moreover, in $V^+$, and for suitable mass parameters, the diagonal components of $S$ 
 each acquire 
the same vacuum expectation value  $\langle S_{\ell\ell}\rangle\neq 0$ and the ground state is flavor universal.
In $V^-$ a finite vacuum expectation value  $\langle S_{\ell\ell}\rangle\neq 0$  arises only for one flavor direction giving rise to a flavorful vacuum. 

\begin{figure}[b]
	\centering
	\begin{minipage}{0.48\columnwidth}
\includegraphics[scale=0.85]{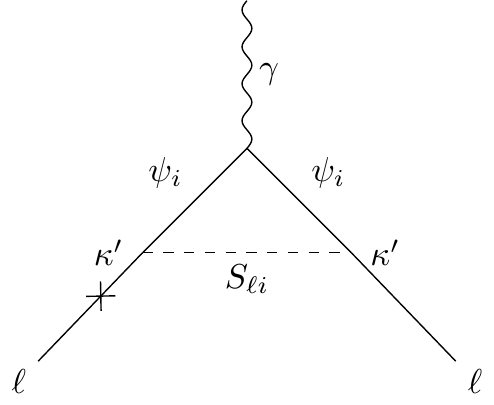}\\
		\centering\footnotesize{$a)$}
\end{minipage}\hspace{0.3em}
\begin{minipage}{0.48\columnwidth}
\includegraphics[scale=0.85]{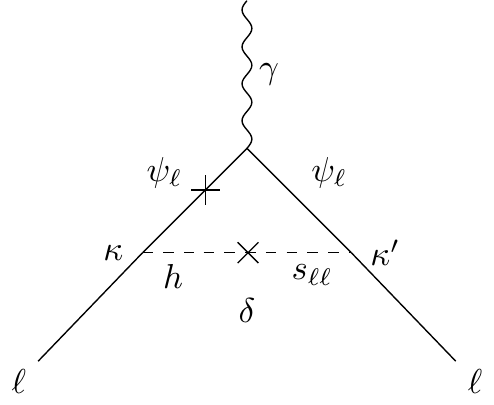}\\
			\centering\footnotesize{$b)$}
\end{minipage}	\caption{Leading loop contributions to   $\Delta a_\ell$ $(\ell=e,\mu,\tau)$, including 
	 {$a)$}  BSM scalar-fermion-loops    with a lepton chiral flip (cross on solid line), and {$b)$} chirally enhanced contributions through scalar mixing (cross on dashed line) provided  $\langle S_{\ell\ell}\rangle \neq 0$, and a BSM fermion $\psi_\ell$ chiral flip (cross on solid line).}
	\label{fig:gminus2feyn}
\end{figure}

\begin{figure}[t]
\begin{center}
	\includegraphics[width=.95\columnwidth]{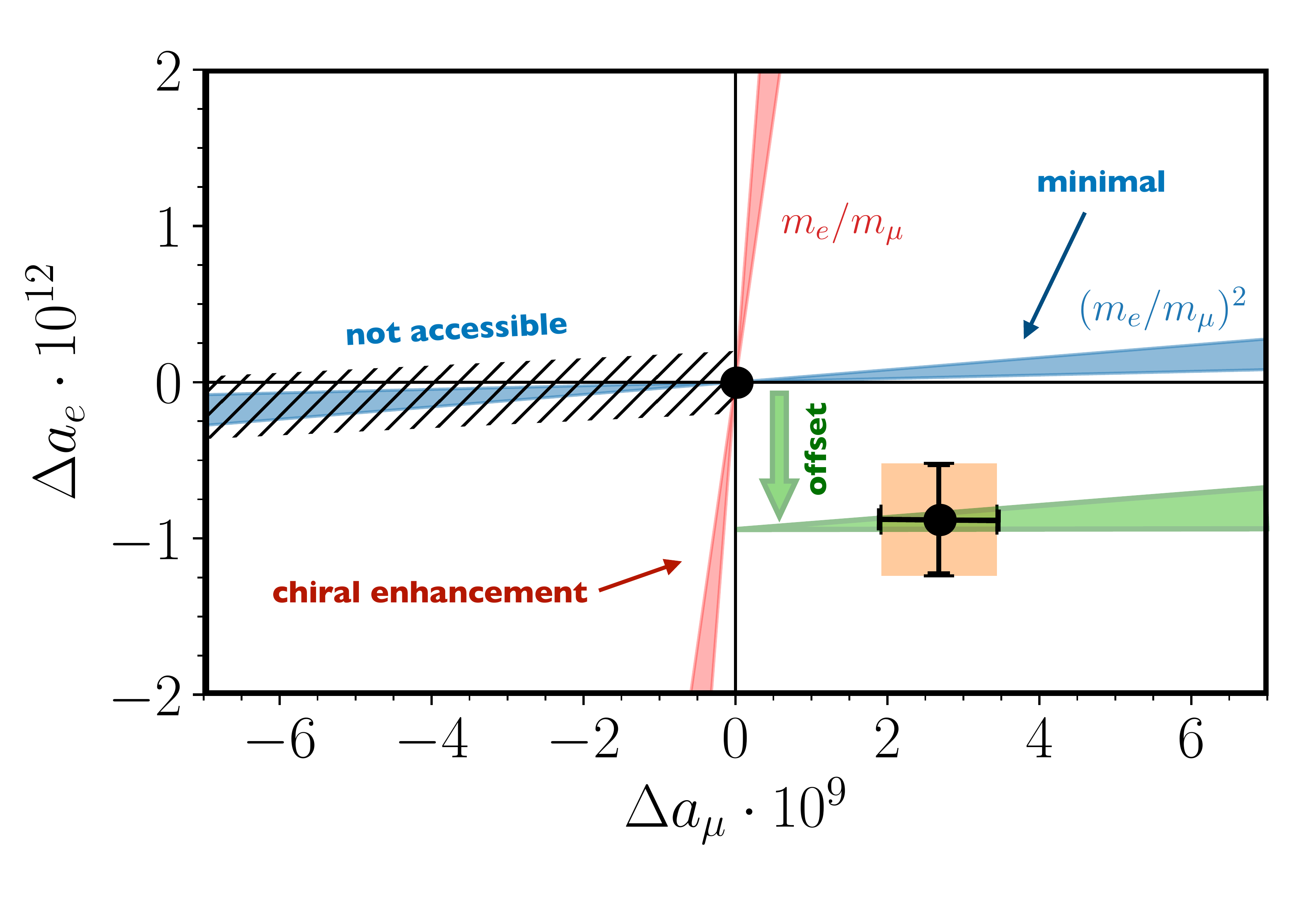}
	\vskip-.5cm
	\caption{Leading  contributions to $\Delta a_{e,\mu}$ 
	from  
	Fig.~\ref{fig:gminus2feyn}{a} 
	    (blue band) and Fig.~\ref{fig:gminus2feyn}{b} (red band), which, 
	     in combination
	    (green band),  explain the electron and muon data  (cross) simultaneously. The chirally enhanced offset is either flavor universal 
	    or points  in the electron direction (green arrow). Band widths  are indicative of a $20\%$ mass splitting between  fermion flavors from  leading loops; the hatched region is inaccessible. }
	    	\vskip-.5cm
	\label{fig:data}
\end{center}
\end{figure}

{\it Explaining anomalous magnetic moments.---} We are now in a position  to explain the data \eq{eq:deltagminus2} in SM extensions  with \eqref{eq:Yukawa} and \eqref{eq:scalarV}. The relevant leading loop effects due to the couplings $\kappa$, $\kappa'$, and $\delta$ are shown in Fig.~\ref{fig:gminus2feyn}, also using  $S=\langle S\rangle +s$. 
 Any lepton  flavor $\ell=e,\mu,\tau$  receives a  contribution from  BSM scalar-fermion loops with chiral flip on the lepton line induced by the coupling $\kappa'$ (see Fig.~\ref{fig:gminus2feyn}{a}). It  scales quadratically with the lepton mass,
\begin{align} \label{eq:mu}
\Delta a_{\ell} =\frac{N_F\,\kappa'^2}{96\pi^2} \, \frac{m_{\ell}^2}{M_{F}^2} \,  f_1\left(\frac{M_S^2}{M_F^2}\right)\,, 
\end{align}
and represents a minimal   lepton flavor dependence with  $f_1(t)  = (2 t^3+3 t^2-6 t^2 \ln t-6 t+1)/(t-1)^4$ positive for any $t$, 
and $f_1(0)=1$.  This manifestly positive contribution is the dominant one for $a_\mu$. 
Contributions through $Z$- and $W$-loops 
are parametrically suppressed   
as
${\cal O}(g_2^2)$ and by fermion mixing \cite{Hiller:2020fbu}.  Comparing \eqref{eq:mu} with the muon   data  for  small  scalar-to-fermion mass ratio $M_S^2/M_F^2\ll 1$ yields the Yukawa coupling $\alpha_{\kappa'}$ 
within $(0.48 \pm 0.15 )(\frac{M_F}{\rm TeV})^{2}$,
which is  large for  TeV-range fermion masses $M_F$. 
Fixing $\Delta a_{\mu}$  to the muon data  (\ref{eq:deltagminus2}) confirms that
the corresponding contribution  \eqref{eq:mu} for the electron  would come out too small and with the wrong sign $\Delta a_e \simeq 6\cdot 10^{-14}$  (see Fig.~\ref{fig:data}).

Additionally, chirally enhanced contributions, which are linear in the lepton mass, may arise
through a portal-mediated scalar mixing where the chiral flip is shifted to a $\psi$ line  (Fig.~\ref{fig:gminus2feyn}{b}).
 The key observation  is that chiral enhancement naturally explains the electron data (Fig.~\ref{fig:data}). In practice, this can be  realized with either $V^+$ or $V^-$. 
If the ground state is $V^-$, it must point into  the electron direction (only $\langle S_{ee}\rangle \neq 0$) or else   \eqref{eq:deltagminus2} cannot be satisfied. 
Overall, this leads to
\begin{equation}\label{eq:ae}
\Delta a_e= 
  \frac{m_{e}}{M_F}\, \frac{\kappa\,\kappa'\, \sin 2\beta }{32\pi^2}  \left[
 f_2(\frac{m_s^2}{M_F^2})-f_2(\frac{m_h^2}{M_F^2})\right]
+\frac{m_e^2}{m_\mu^2} \Delta a_{\mu}
\end{equation}
where $m_{h,s}$ are the Higgs and the BSM scalar mass, and the last term accounts for \eq{eq:mu}.
The loop function $f_2(t)= (3 t^2-2 t^2 \ln t-4 t+1)/(1-t)^3$ is  positive for any $t$ and $f_2(0)=1$. The  mixing angle $\beta$ between the scalar $s_{\ell\ell}$ and  the physical  Higgs $h$ is fixed via 
\begin{equation} \label{eq:s2b}
\begin{aligned}
\tan 2\beta = \frac{\delta}{\sqrt{\lambda(u+v)}} \frac{m_h}{m_s}\bigg(1+  \mathcal{O}( m_h^2/m_s^2)\bigg)\,.
\end{aligned}
\end{equation}
In \eqref{eq:ae}, the 
 term linear 
 in the electron mass provides a unique
offset for the electron $\Delta a_e$,  sketched in Fig.~\ref{fig:data}. It dominates parametrically over the quadratic term
and can have either sign 
set by the Yukawas $\kappa$, $\kappa'$ 
and the portal coupling $\delta$.

\begin{figure}[b]
	\includegraphics[width=.9\columnwidth]{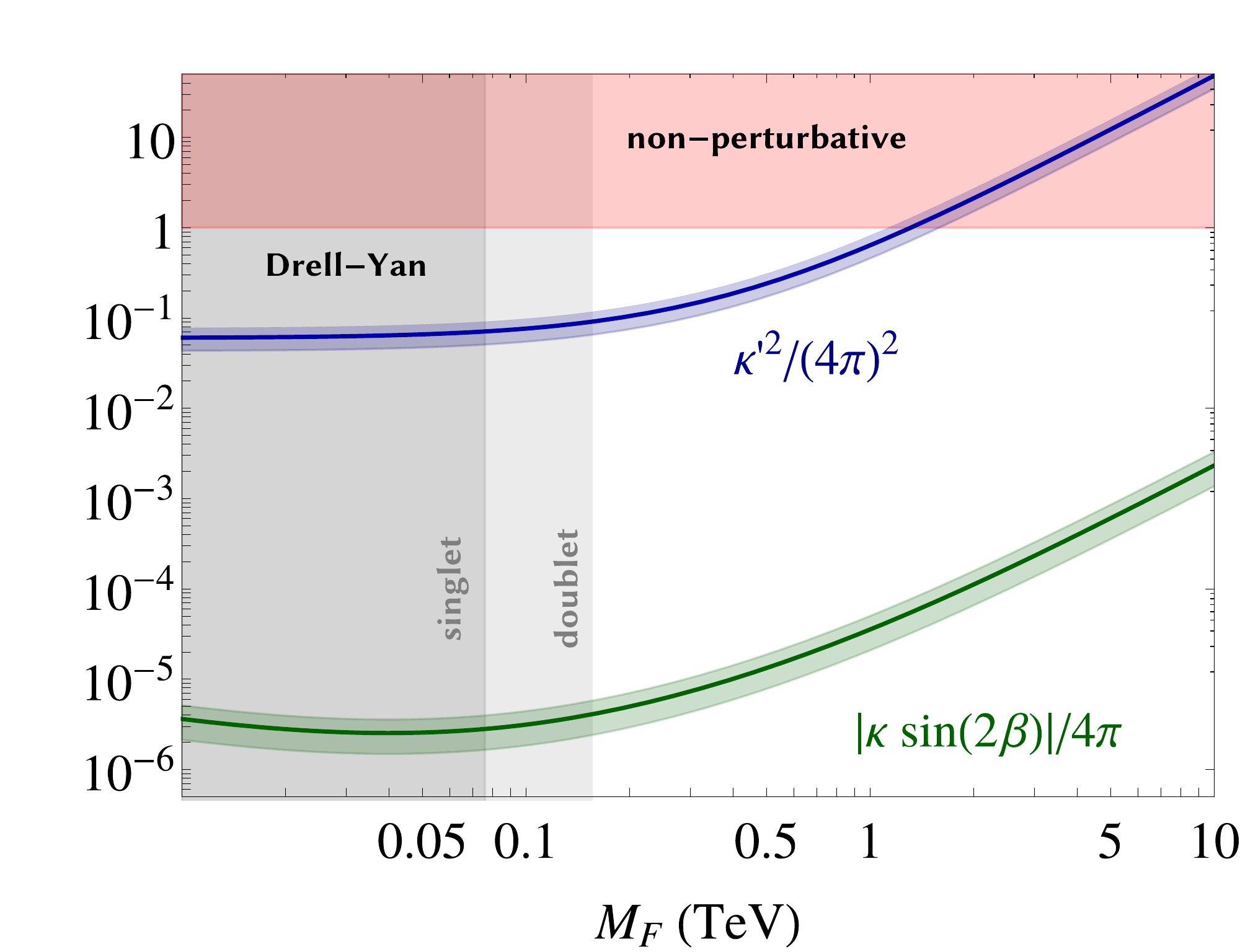}
	\vskip-.2cm
	\caption{Window for Yukawa and portal couplings which simultaneously explain  the muon and electron data (\ref{eq:deltagminus2}) as functions of the BSM fermion mass $M_F$ ($M_S = 0.5$ TeV, band width $1\sigma$). 	Gray-shaded areas are excluded by Drell-Yan searches,  the red-shaded area indicates  strong coupling. All results refer to $V^-$, very similar ones  are found for $V^+$ (not shown).}
	\label{fig:1}
	\vskip-.3cm
\end{figure}
\begin{figure*}
	\vskip-.5cm
	\includegraphics[width=1.8\columnwidth]{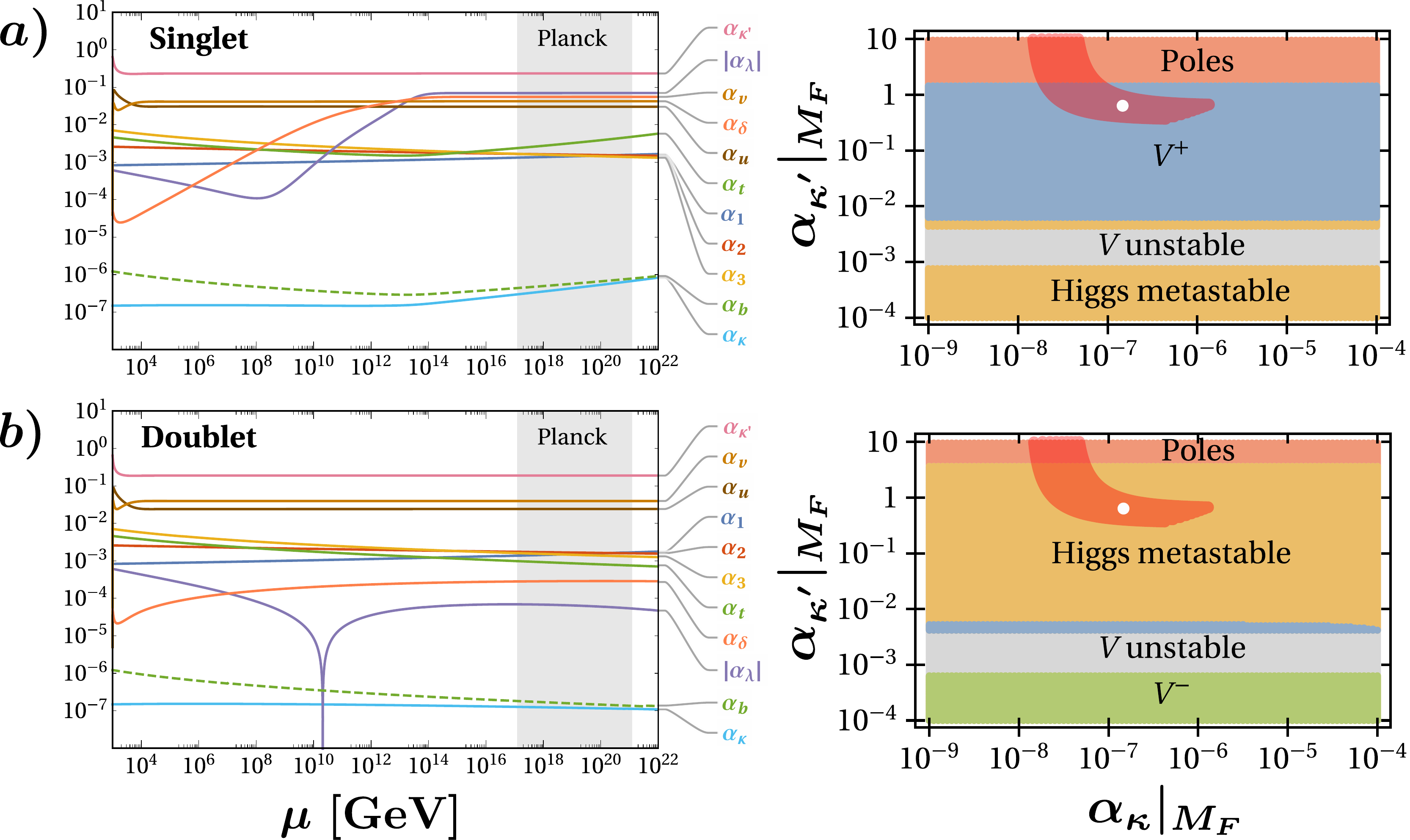}
	\vskip-.3cm
	\caption{Benchmark trajectories  $(M_F=2M_S=1$~TeV)  between the matching scale $M_F$ and the Planck scale (left), and parameter scans of vacua at the Planck scale  (right) for  $a)$ the  singlet model (top) and $b)$ the doublet model (bottom) using
 $(\alpha_\delta,\alpha_u,\alpha_v,\alpha_y)|_{M_F}=(5,-1, 4,0)\cdot 10^{-5}$.  
High scale  vacua are shown as functions
 of  the Yukawa couplings $(\alpha_\kappa,\alpha_\kappa')|_{M_F}$. 
 Parameters within   the red-shaded areas are
compatible with    data \eq{eq:deltagminus2}; 
white dots refer to the benchmarks  on the left.}
			\label{fig:combined}
\end{figure*}
As an estimate, comparing \eqref{eq:ae} with the electron  data  assuming 
$m_h^2/M_F^2\ll 1$ and simultaneously fixing \eqref{eq:mu}
to match the muon data, we find  
$|{\kappa}\sin 2\beta| \simeq (2.9 \pm 1.2) \cdot 10^{-4}(\frac{M_F}{\rm TeV})^2\,.$
The full parameter window explaining the data is indicated in Fig.~\ref{fig:1}  assuming $V^-$. 
Corrections from $Z$- and $W$-exchange, which contribute differently in the singlet and doublet models, are suppressed by small fermion mixing angles and not sizeable enough to be seen in Fig.~\ref{fig:1}. Also shown are limits on $M_F$ (gray) from Drell-Yan processes \cite{Alves:2014cda,Farina:2016rws,Hiller:2019tvg} and on perturbativity in   $\alpha_{\kappa'}$ (red). 
We observe
$M_F$ within $(0.05-2)$~TeV for  $\alpha_{\kappa'}$ within $(10^{-2}-1)$, with
$\kappa\sin 2\beta/(4\pi) $ deeply perturbative (green) 
 for  small portal coupling $\delta$.  The dual parameter space  $(\kappa'\ll \kappa)$  where Fig.~\ref{fig:gminus2feyn}a is replaced by the corresponding Higgs-fermion loops, is ruled out by $Z\to\ell\ell$ data \cite{Tanabashi:2018oca}, 
 which constrains left-handed (right-handed) fermion mixing angles in the singlet (doublet) model to be of $\mathcal{O}(10^{-2})$ or smaller.

 If the vacuum  is $V^+$, {all}  lepton anomalous magnetic moments  receive a chirally enhanced contribution from Fig.~\ref{fig:gminus2feyn}{b}, similar to the first term in \eq{eq:ae}. The  offset in Fig.~\ref{fig:data} is then slightly tilted and points  along the direction of the red band. Due to the smallness of the tilt, results and  constraints are   similar to those  for $V^-$ in Fig.~\ref{fig:1}.

{\it Running of couplings up to the Planck scale.---} We now turn to the RG running of couplings and conditions under which  models   
are  stable and predictive  up to the Planck scale.
We normalize couplings  to   loop factors,
\begin{equation}\label{couplings-gauge}
\alpha_x=\frac{x^2}{(4\pi)^2}\,,\quad 
\alpha_z=\frac{z}{(4\pi)^2}\,, 
\end{equation}
where $x=g_1, g_2, g_3, y_t, y_b, y, \kappa, \kappa'$ are any of the gauge, top, bottom or BSM Yukawa couplings, and $z=\lambda, u,v, \delta$ are the quartic and portal couplings. Models are matched onto the SM at the  scale set by the fermion mass.  
For the running above $M_F$, we retain all 12 RG beta-functions up to  two-loop order in all couplings \cite{Machacek:1983tz,Machacek:1983fi,Machacek:1984zw,Luo:2002ti}.

The left panel of Fig.~\ref{fig:combined} shows  benchmark trajectories   up to the Planck scale  $M_{\rm Pl}$ for models  starting in  the vacuum  $V^-$  at the scale $M_F$.
For some initial conditions  $\alpha_{{}_{\rm BSM}}|_{M_F}$  at the low scale,  such as those used in Fig.~\ref{fig:combined}, we find that the running is stable up to the Planck scale. 
We also observe from Fig.~\ref{fig:combined} that the Higgs potential becomes stable (remains metastable) in the singlet (doublet) model. 
Higgs stability in the doublet model   
can be achieved for 
 larger portal and quartic couplings. 
Some couplings in Fig.~\ref{fig:combined}  run slowly all the way up to the Planck scale. Others show a slow or fast  cross-over to 
 near-constant values
due  to near-zeros of  beta functions    \cite{Holdom:1984sk}
  which arise from a competition between SM and BSM matter. 
In the absence of quantum gravity, the evolution of couplings
ultimately  terminates in  an  interacting UV fixed point  
 corresponding to asymptotic safety   (singlet benchmark) with asymptotic freedom prevailing in the weak and strong sectors  \cite{Bond:2016dvk,Bond:2017wut,Kowalska:2017fzw}. In some cases, 
trajectories remain safe up to the Planck scale (doublet benchmark) but  blow up at transplanckian energies. 
For other initial conditions we also find unsafe trajectories which terminate in subplanckian Landau poles 
(see \cite{Hiller:2020fbu} for a detailed study of  initial conditions $\alpha_{{}_{\rm BSM}}|_{M_F}$).

The right panel of Fig.~\ref{fig:combined}  shows  the vacua  of singlet and doublet models  at the Planck scale in terms of the   Yukawa couplings $(\alpha_\kappa,\alpha_{\kappa'})|_{M_F}$  at the matching scale.	Integrating the RG between $M_F$ and $M_{\rm Pl}$,  we find wide ranges of  models whose vacua at the Planck scale are either $V^+$ (blue), or a stable $V$ with a metastable Higgs sector ($\alpha_\lambda\gtrsim -10^{-4}$) such as in the SM  \cite{Degrassi:2012ry,Buttazzo:2013uya} (yellow). For other parameter ranges we also find $V^-$ (green), or unstable BSM potentials  (gray), or  Landau poles below the Planck scale (light red). Most importantly, the anomalous magnetic moments \eq{eq:deltagminus2} are matched  for  couplings in the  red-shaded areas which cover the $1\sigma$ band.  Constraints from Higgs signal strength    \cite{Tanabashi:2018oca} imply  an upper bound on $\alpha_\kappa$ corresponding to  a lower bound for the scalar mass  of about 226~GeV (for $M_F=1$~TeV). Similar results   are found for $V^+$ at the low scale (not shown) except that  regions with $V^-$  in Fig.~\ref{fig:combined}   turn into $V^+$. We conclude that  models are stable and Planck-safe for a  range of parameters $\alpha_{{}_{\rm BSM}}|_{M_F}$.

{\it Collider production and decay.---}  Models predict new scalars and fermions in the TeV  energy range. Their phenomenology  is characterized  by an enlarged flavor sector with a large Yukawa coupling $\kappa'$ and  moderate or small  couplings $\kappa$, $\delta$.
We   identify collider signatures through production and decay 
 \cite{Hiller:2020fbu}. We denote the fermions in the singlet model by $\psi^{-1}_s$ and the 
   isospin components in the doublet model by $\psi^0_d$ and $\psi_d^{-1}$; superscripts show electric charge.
The $\psi^0_d$ is   lighter than the $\psi^{-1}_d$ by $\Delta m=  M_{\psi^{-1}} -M_{\psi^{0}}=g_2^2\, \sin \theta_W^2 \,m_Z/(8 \pi)     \simeq 0.4$GeV  \cite{Cirelli:2005uq}.
All fermion flavors can be pair-produced in  $pp$ and $\ell\ell$ machines  via $s$-channel $\gamma$ or $Z$ exchange, and  through $W^\pm$ exchange at $pp$-colliders (doublet model only).
Lepton colliders allow for pair-production from $t$-channel $S$ at order $\kappa'^2$, which is  sizable (see Fig.~\ref{fig:1}).
Single $\psi$ production together with a lepton arises from $s$-channel $Z$- and $W$-boson contributions via fermion mixing. $S$ production occurs only via  the  Higgs portal, or at lepton colliders  with $t$-channel $\psi$ in association with $h$ at order $\kappa\, \kappa^\prime$
or in pairs at order $(\kappa')^2$.

If kinematically allowed, the charged  fermions decay as $\psi^{-1} \to S \ell $ and the neutral ones as $\psi^0_d\to S \nu$. If these channels are closed, the 
$\psi^{-1}$
decay to 
Higgs plus lepton instead. The decay rate
$\Gamma(\psi^{-1}\rightarrow h\,\ell^-) = \frac{\kappa^2}{64\pi}{ M_F}(1-{m_h^2}/{M_F^2})^2$
provides the lifetime estimate    $\Gamma^{-1} \sim 10^{-27} (1/\alpha_{\kappa})(1/M_F [{\rm TeV}])\,{\rm s}$.
The neutral  fermion $\psi_d^0$
cascades down  slower, yet still promptly through $W$-emission with $\psi^{0}_{d,i} \to  \psi_{d,i}^{-1 *} W^{+*} \to h \ell_i^- W^{+*}$.
If kinematically allowed, the BSM scalars $S$ undergo tree level decay into $\psi \bar \psi$ via $y$, and into
$\psi \,\ell$ via $\kappa^\prime$. At one-loop arise the decays  $S \to \gamma \gamma$, $Z Z$, $Z \gamma$, and $ S\to  WW$ (doublet model only)
 from $y$. Although there is no  genuine lepton flavor violation (LFV)
 as  flavor in the $S$-decay process is conserved, the mixing between the  $\psi$ and the SM leptons introduces very distinct LFV-like final states
$S_{ij} \to \ell_i^\pm \ell_j^\mp$.  The LFV-like decays at the order $\kappa \kappa' v_h/M_F$   or     $(\kappa^\prime)^2 (v_s v_h/2 M_F)^2$ 
 are the leading ones   
 for negligible $y$ and $M_S/M_F\ll 1$.

{\it Discussion.---} We have shown that extensions of the standard model with new vector-like leptons and singlet matrix scalar fields  \eqref{eq:Yukawa}, \eqref{eq:scalarV} explain the muon and electron  anomalous magnetic moments  \eq{eq:deltagminus2}
simultaneously. 
Yukawa couplings mixing SM and BSM matter and a Higgs portal coupling are instrumental 
to generate both  minimal \eq{eq:mu} and chirally enhanced \eq{eq:ae} contributions, which, when taken together, match the present data   (Fig.~\ref{fig:data}).
Also, the  mechanism generating anomalous magnetic moments is  rather natural and not fine-tuned to the data. In fact, 
our models can in principle accommodate
deviations $\Delta a_\mu$  and  $\Delta a_e$   in the half-plane spanned by the minimal and chirally enhanced contributions as indicated in Fig.~\ref{fig:data}.

 Further features   unlike in the SM 
 are a stable  Higgs  potential, and well-behaved running couplings  up to the Planck scale.
This includes  asymptotically safe extensions of the SM
which, for the first time, match the measured values of all gauge couplings and the Higgs, top and bottom masses,
and models which may run into  poles or instabilities at  transplanckian energies.
Also, some parameter settings can explain the data but are unsafe at high energies due to poles prior to the Planck scale   (Fig.~\ref{fig:combined}). We thus see very clearly how the high energy behavior offers an additional selection criterion for models and their low energy BSM parameters.
Further predictions are a strongly and a weakly coupled Yukawa sector, and new matter fields with masses  in the TeV  range (Fig.~\ref{fig:1}) which can be tested at colliders.

From the viewpoint of lepton universality, it is worth noting that a manifest breaking
has been instrumental
in all previous models explaining both anomalies \eq{eq:deltagminus2}.
As a proof of principle, however,
our models find   that any  breaking beyond SM Yukawas
is not mandatory. 
In a related vein, 
we also stress that lepton universality in 
itself is not key for  asymptotic safety. In fact, 
it would be straightforward to explicitly break lepton flavor universality in  alterations of  models while maintaining predictions for both anomalies,
and  without spoiling 
a well-behaved high energy behavior.

Another  aspect which sets our models apart from any  previous ones explaining both anomalies 
is that we also predict the deviation of the tau anomalous magnetic moment from its standard model  value. This can be done   solely using the data  \eq{eq:deltagminus2} and the  vacuum, and is insensitive to  any other details. Specifically,
provided the ground state  distinguishes  electron flavor we have
\beq
\Delta a_\tau \equiv a^{\rm exp}_\tau-a^{\rm SM}_\tau=(7.5\pm 2.1)\cdot 10^{-7}\,,
\eeq
and
$\Delta a_\tau=(8.1\pm  2.2)\cdot 10^{-7}$
otherwise. Although the present limit on $\Delta a_\tau$ is four orders of magnitude away  \cite{Tanabashi:2018oca}, it would be very interesting to test these
predictions  in the future. We also note that with small CP phases,  the electric dipole moment of the electron can be as large as the present bound $d_e < 1.1 \cdot 10^{-29}$~ecm \cite{Andreev:2018ayy}. In settings with flavor universal vacua the bound extends to  all lepton electric dipole moments  $d_\ell$, which would make an experimental check for the muon and the tau very challenging.

Finally, we comment on  asymptotic safety as a guiding principle for model building. 
Vector-like fermions alongside singlet matrix scalar fields and their Yukawa interactions are   established ingredients in settings with perturbatively exact asymptotic safety, and appear  prominently in templates for asymptotically safe SM extensions. 
Here, we have extended earlier ideas   by additionally allowing for new Yukawa and portal interactions between SM and BSM matter. Curiously, these  new interactions   not only improve the high energy 
behavior    (Fig.~\ref{fig:combined}) in the spirit of asymptotic safety, 
but also generate anomalous magnetic moments  (Fig.~\ref{fig:gminus2feyn}) which can match the data naturally  (Fig.~\ref{fig:data}). 
It would thus seem interesting to further explore the potential of models inspired by asymptotic safety for flavor and particle  physics. 
\\[1ex]

{\it Acknowledgements.---}
This work is partly supported  by the DFG Research Unit FOR 1873 ``Quark Flavour Physics and Effective Field Theories''. GH and DL  thank the SLAC Theory Group for hospitality during the final stages of this work.\\[1ex]

{\it Note added.---} 
The  possibility of rendering $ \Delta a_\mu$  insignificant has recently been suggested by a  lattice determination of the hadronic vacuum polarization \cite{Borsanyi:2020mff}. Note, though, that these findings are in tension with electroweak data \cite{Crivellin:2020zul,Keshavarzi:2020bfy} and other lattice studies, which requires further scrutiny \cite{Aoyama:2020ynm}. 

\bibliographystyle{JHEP}
\bibliography{hhls-AMM}

\providecommand{\href}[2]{#2}\begingroup\raggedright\begin{thebibliography}{10}

\bibitem{Tanabashi:2018oca}
{\scshape Particle Data Group} collaboration, \emph{{Review of Particle
  Physics}}, \href{https://doi.org/10.1103/PhysRevD.98.030001}{\emph{Phys.
  Rev.} {\bfseries D98} (2018) 030001}.

\bibitem{Hanneke:2008tm}
D.~Hanneke, S.~Fogwell and G.~Gabrielse, \emph{{New Measurement of the Electron
  Magnetic Moment and the Fine Structure Constant}},
  \href{https://doi.org/10.1103/PhysRevLett.100.120801}{\emph{Phys. Rev. Lett.}
  {\bfseries 100} (2008) 120801}
  [\href{https://arxiv.org/abs/0801.1134}{{\ttfamily 0801.1134}}].

\bibitem{Parker:2018vye}
R.~H. Parker, C.~Yu, W.~Zhong, B.~Estey and H.~Müller, \emph{{Measurement of
  the fine-structure constant as a test of the Standard Model}},
  \href{https://doi.org/10.1126/science.aap7706}{\emph{Science} {\bfseries 360}
  (2018) 191} [\href{https://arxiv.org/abs/1812.04130}{{\ttfamily
  1812.04130}}].

\bibitem{Jegerlehner:2017lbd}
F.~Jegerlehner, \emph{{Muon g – 2 theory: The hadronic part}},
  \href{https://doi.org/10.1051/epjconf/201816600022}{\emph{EPJ Web Conf.}
  {\bfseries 166} (2018) 00022}
  [\href{https://arxiv.org/abs/1705.00263}{{\ttfamily 1705.00263}}].

\bibitem{Davier:2016iru}
M.~Davier, \emph{{Update of the Hadronic Vacuum Polarisation Contribution to
  the muon g-2}},
  \href{https://doi.org/10.1016/j.nuclphysbps.2017.03.047}{\emph{Nucl. Part.
  Phys. Proc.} {\bfseries 287-288} (2017) 70}
  [\href{https://arxiv.org/abs/1612.02743}{{\ttfamily 1612.02743}}].

\bibitem{Davoudiasl:2018fbb}
H.~Davoudiasl and W.~J. Marciano, \emph{{Tale of two anomalies}},
  \href{https://doi.org/10.1103/PhysRevD.98.075011}{\emph{Phys. Rev.}
  {\bfseries D98} (2018) 075011}
  [\href{https://arxiv.org/abs/1806.10252}{{\ttfamily 1806.10252}}].

\bibitem{Liu:2018xkx}
J.~Liu, C.~E.~M. Wagner and X.-P. Wang, \emph{{A light complex scalar for the
  electron and muon anomalous magnetic moments}},
  \href{https://doi.org/10.1007/JHEP03(2019)008}{\emph{JHEP} {\bfseries 03}
  (2019) 008} [\href{https://arxiv.org/abs/1810.11028}{{\ttfamily
  1810.11028}}].

\bibitem{Gardner:2019mcl}
S.~Gardner and X.~Yan, \emph{{Light scalars with lepton number to solve the
  $(g-2)_e$ anomaly}},  \href{https://arxiv.org/abs/1907.12571}{{\ttfamily
  1907.12571}}.

\bibitem{Bauer:2019gfk}
M.~Bauer, M.~Neubert, S.~Renner, M.~Schnubel and A.~Thamm, \emph{{Axionlike
  Particles, Lepton-Flavor Violation, and a New Explanation of $a_\mu$ and
  $a_e$}}, \href{https://doi.org/10.1103/PhysRevLett.124.211803}{\emph{Phys.
  Rev. Lett.} {\bfseries 124} (2020) 211803}
  [\href{https://arxiv.org/abs/1908.00008}{{\ttfamily 1908.00008}}].

\bibitem{Dutta:2018fge}
B.~Dutta and Y.~Mimura, \emph{{Electron $g-2$ with flavor violation in MSSM}},
  \href{https://doi.org/10.1016/j.physletb.2018.12.070}{\emph{Phys. Lett.}
  {\bfseries B790} (2019) 563}
  [\href{https://arxiv.org/abs/1811.10209}{{\ttfamily 1811.10209}}].

\bibitem{Endo:2019bcj}
M.~Endo and W.~Yin, \emph{{Explaining electron and muon $g-2$ anomaly in SUSY
  without lepton-flavor mixings}},
  \href{https://arxiv.org/abs/1906.08768}{{\ttfamily 1906.08768}}.

\bibitem{Badziak:2019gaf}
M.~Badziak and K.~Sakurai, \emph{{Explanation of electron and muon g-2
  anomalies in the MSSM}},  \href{https://arxiv.org/abs/1908.03607}{{\ttfamily
  1908.03607}}.

\bibitem{Crivellin:2018qmi}
A.~Crivellin, M.~Hoferichter and P.~Schmidt-Wellenburg, \emph{{Combined
  explanations of $(g-2)_{\mu,e}$ and implications for a large muon EDM}},
  \href{https://doi.org/10.1103/PhysRevD.98.113002}{\emph{Phys. Rev.}
  {\bfseries D98} (2018) 113002}
  [\href{https://arxiv.org/abs/1807.11484}{{\ttfamily 1807.11484}}].

\bibitem{Crivellin:2019mvj}
A.~Crivellin and M.~Hoferichter, \emph{{Combined explanations of $(g-2)_\mu$,
  $(g-2)_e$ and implications for a large muon EDM}},  in \emph{{33rd Rencontres
  de Physique de La Vallée d'Aoste (LaThuile 2019) La Thuile, Aosta, Italy,
  March 10-16, 2019}}, 2019, \href{https://arxiv.org/abs/1905.03789}{{\ttfamily
  1905.03789}}.

\bibitem{Han:2018znu}
X.-F. Han, T.~Li, L.~Wang and Y.~Zhang, \emph{{Simple interpretations of lepton
  anomalies in the lepton-specific inert two-Higgs-doublet model}},
  \href{https://doi.org/10.1103/PhysRevD.99.095034}{\emph{Phys. Rev.}
  {\bfseries D99} (2019) 095034}
  [\href{https://arxiv.org/abs/1812.02449}{{\ttfamily 1812.02449}}].

\bibitem{Abdullah:2019ofw}
M.~Abdullah, B.~Dutta, S.~Ghosh and T.~Li, \emph{{$(g-2)_{\mu,e}$ and the ANITA
  anomalous events in a three-loop neutrino mass model}},
  \href{https://doi.org/10.1103/PhysRevD.100.115006}{\emph{Phys. Rev. D}
  {\bfseries 100} (2019) 115006}
  [\href{https://arxiv.org/abs/1907.08109}{{\ttfamily 1907.08109}}].

\bibitem{Bond:2017wut}
A.~D. Bond, G.~Hiller, K.~Kowalska and D.~F. Litim, \emph{{Directions for model
  building from asymptotic safety}},
  \href{https://doi.org/10.1007/JHEP08(2017)004}{\emph{JHEP} {\bfseries 08}
  (2017) 004} [\href{https://arxiv.org/abs/1702.01727}{{\ttfamily
  1702.01727}}].

\bibitem{Kowalska:2017fzw}
K.~Kowalska, A.~Bond, G.~Hiller and D.~Litim, \emph{{Towards an asymptotically
  safe completion of the Standard Model}},
  \href{https://doi.org/10.22323/1.314.0542}{\emph{PoS} {\bfseries EPS-HEP2017}
  (2017) 542}.

\bibitem{Litim:2014uca}
D.~F. Litim and F.~Sannino, \emph{{Asymptotic safety guaranteed}},
  \href{https://doi.org/10.1007/JHEP12(2014)178}{\emph{JHEP} {\bfseries 12}
  (2014) 178} [\href{https://arxiv.org/abs/1406.2337}{{\ttfamily 1406.2337}}].

\bibitem{Wilson:1971bg}
K.~G. Wilson, \emph{{Renormalization group and critical phenomena. 1.
  Renormalization group and the Kadanoff scaling picture}},
  \href{https://doi.org/10.1103/PhysRevB.4.3174}{\emph{Phys.Rev.} {\bfseries
  B4} (1971) 3174}.

\bibitem{Bailin:1974bq}
D.~Bailin and A.~Love, \emph{{Asymptotic Near Freedom}},
  \href{https://doi.org/10.1016/0550-3213(74)90470-2}{\emph{Nucl. Phys.}
  {\bfseries B75} (1974) 159}.

\bibitem{Weinberg:1980gg}
S.~Weinberg, \emph{{Ultraviolet divergences in quantum theories of
  gravitation}}, {\emph{in: General Relativity: An Einstein centenary survey,
  Eds. Hawking, S.W., Israel, W; Cambridge University Press} (1979) 790}.

\bibitem{Bond:2016dvk}
A.~D. Bond and D.~F. Litim, \emph{{Theorems for Asymptotic Safety of Gauge
  Theories}}, \href{https://doi.org/10.1140/epjc/s10052-017-4976-5,
  10.1140/epjc/s10052-017-5034-z}{\emph{Eur. Phys. J.} {\bfseries C77} (2017)
  429} [\href{https://arxiv.org/abs/1608.00519}{{\ttfamily 1608.00519}}].

\bibitem{Bond:2018oco}
A.~D. Bond and D.~F. Litim, \emph{{Price of Asymptotic Safety}},
  \href{https://doi.org/10.1103/PhysRevLett.122.211601}{\emph{Phys. Rev. Lett.}
  {\bfseries 122} (2019) 211601}
  [\href{https://arxiv.org/abs/1801.08527}{{\ttfamily 1801.08527}}].

\bibitem{Bond:2017tbw}
A.~D. Bond, D.~F. Litim, G.~Medina~Vazquez and T.~Steudtner, \emph{{UV
  conformal window for asymptotic safety}},
  \href{https://doi.org/10.1103/PhysRevD.97.036019}{\emph{Phys. Rev.}
  {\bfseries D97} (2018) 036019}
  [\href{https://arxiv.org/abs/1710.07615}{{\ttfamily 1710.07615}}].

\bibitem{Bond:2019npq}
A.~D. Bond, D.~F. Litim and T.~Steudtner, \emph{{Asymptotic safety with
  Majorana fermions and new large $N$ equivalences}},
  \href{https://doi.org/10.1103/PhysRevD.101.045006}{\emph{Phys. Rev. D}
  {\bfseries 101} (2020) 045006}
  [\href{https://arxiv.org/abs/1911.11168}{{\ttfamily 1911.11168}}].

\bibitem{Bond:2017lnq}
A.~D. Bond and D.~F. Litim, \emph{{More asymptotic safety guaranteed}},
  \href{https://doi.org/10.1103/PhysRevD.97.085008}{\emph{Phys. Rev.}
  {\bfseries D97} (2018) 085008}
  [\href{https://arxiv.org/abs/1707.04217}{{\ttfamily 1707.04217}}].

\bibitem{Bond:2017suy}
A.~D. Bond and D.~F. Litim, \emph{{Asymptotic safety guaranteed in
  supersymmetry}},
  \href{https://doi.org/10.1103/PhysRevLett.119.211601}{\emph{Phys. Rev. Lett.}
  {\bfseries 119} (2017) 211601}
  [\href{https://arxiv.org/abs/1709.06953}{{\ttfamily 1709.06953}}].

\bibitem{Barducci:2018ysr}
D.~Barducci, M.~Fabbrichesi, C.~M. Nieto, R.~Percacci and V.~Skrinjar,
  \emph{{In search of a UV completion of the standard model — 378,000 models
  that don’t work}},
  \href{https://doi.org/10.1007/JHEP11(2018)057}{\emph{JHEP} {\bfseries 11}
  (2018) 057} [\href{https://arxiv.org/abs/1807.05584}{{\ttfamily
  1807.05584}}].

\bibitem{Hiller:2019tvg}
G.~Hiller, C.~Hormigos-Feliu, D.~F. Litim and T.~Steudtner,
  \emph{{Asymptotically safe extensions of the Standard Model with flavour
  phenomenology}},  in \emph{{54th Rencontres de Moriond on Electroweak
  Interactions and Unified Theories (Moriond EW 2019) La Thuile, Italy, March
  16-23, 2019}}, 2019, \href{https://arxiv.org/abs/1905.11020}{{\ttfamily
  1905.11020}}.

\bibitem{Hiller:2020fbu}
G.~Hiller, C.~Hormigos-Feliu, D.~F. Litim and T.~Steudtner, \emph{{Model
  Building from Asymptotic Safety with Higgs and Flavor Portals}},
  {\emph{{\upshape to appear in} Phys. Rev. D} (2020) }
  [\href{https://arxiv.org/abs/2008.08606}{{\ttfamily 2008.08606}}].

\bibitem{Alves:2014cda}
D.~S.~M. Alves, J.~Galloway, J.~T. Ruderman and J.~R. Walsh, \emph{{Running
  Electroweak Couplings as a Probe of New Physics}},
  \href{https://doi.org/10.1007/JHEP02(2015)007}{\emph{JHEP} {\bfseries 02}
  (2015) 007} [\href{https://arxiv.org/abs/1410.6810}{{\ttfamily 1410.6810}}].

\bibitem{Farina:2016rws}
M.~Farina, G.~Panico, D.~Pappadopulo, J.~T. Ruderman, R.~Torre and A.~Wulzer,
  \emph{{Energy helps accuracy: electroweak precision tests at hadron
  colliders}},
  \href{https://doi.org/10.1016/j.physletb.2017.06.043}{\emph{Phys. Lett.}
  {\bfseries B772} (2017) 210}
  [\href{https://arxiv.org/abs/1609.08157}{{\ttfamily 1609.08157}}].

\bibitem{Machacek:1983tz}
M.~E. Machacek and M.~T. Vaughn, \emph{{Two Loop Renormalization Group
  Equations in a General Quantum Field Theory. 1. Wave Function
  Renormalization}},
  \href{https://doi.org/10.1016/0550-3213(83)90610-7}{\emph{Nucl. Phys.}
  {\bfseries B222} (1983) 83}.

\bibitem{Machacek:1983fi}
M.~E. Machacek and M.~T. Vaughn, \emph{{Two Loop Renormalization Group
  Equations in a General Quantum Field Theory. 2. Yukawa Couplings}},
  \href{https://doi.org/10.1016/0550-3213(84)90533-9}{\emph{Nucl. Phys.}
  {\bfseries B236} (1984) 221}.

\bibitem{Machacek:1984zw}
M.~E. Machacek and M.~T. Vaughn, \emph{{Two Loop Renormalization Group
  Equations in a General Quantum Field Theory. 3. Scalar Quartic Couplings}},
  \href{https://doi.org/10.1016/0550-3213(85)90040-9}{\emph{Nucl. Phys.}
  {\bfseries B249} (1985) 70}.

\bibitem{Luo:2002ti}
M.-x. Luo, H.-w. Wang and Y.~Xiao, \emph{{Two loop renormalization group
  equations in general gauge field theories}},
  \href{https://doi.org/10.1103/PhysRevD.67.065019}{\emph{Phys. Rev.}
  {\bfseries D67} (2003) 065019}
  [\href{https://arxiv.org/abs/hep-ph/0211440}{{\ttfamily hep-ph/0211440}}].

\bibitem{Holdom:1984sk}
B.~Holdom, \emph{{Techniodor}},
  \href{https://doi.org/10.1016/0370-2693(85)91015-9}{\emph{Phys. Lett.}
  {\bfseries 150B} (1985) 301}.

\bibitem{Degrassi:2012ry}
G.~Degrassi, S.~Di~Vita, J.~Elias-Miro, J.~R. Espinosa, G.~F. Giudice,
  G.~Isidori et~al., \emph{{Higgs mass and vacuum stability in the Standard
  Model at NNLO}}, \href{https://doi.org/10.1007/JHEP08(2012)098}{\emph{JHEP}
  {\bfseries 08} (2012) 098} [\href{https://arxiv.org/abs/1205.6497}{{\ttfamily
  1205.6497}}].

\bibitem{Buttazzo:2013uya}
D.~Buttazzo, G.~Degrassi, P.~P. Giardino, G.~F. Giudice, F.~Sala, A.~Salvio
  et~al., \emph{{Investigating the near-criticality of the Higgs boson}},
  \href{https://doi.org/10.1007/JHEP12(2013)089}{\emph{JHEP} {\bfseries 12}
  (2013) 089} [\href{https://arxiv.org/abs/1307.3536}{{\ttfamily 1307.3536}}].

\bibitem{Cirelli:2005uq}
M.~Cirelli, N.~Fornengo and A.~Strumia, \emph{{Minimal dark matter}},
  \href{https://doi.org/10.1016/j.nuclphysb.2006.07.012}{\emph{Nucl. Phys.}
  {\bfseries B753} (2006) 178}
  [\href{https://arxiv.org/abs/hep-ph/0512090}{{\ttfamily hep-ph/0512090}}].

\bibitem{Andreev:2018ayy}
{\scshape ACME} collaboration, \emph{{Improved limit on the electric dipole
  moment of the electron}},
  \href{https://doi.org/10.1038/s41586-018-0599-8}{\emph{Nature} {\bfseries
  562} (2018) 355}.

\bibitem{Borsanyi:2020mff}
S.~Borsanyi et~al., \emph{{Leading-order hadronic vacuum polarization
  contribution to the muon magnetic momentfrom lattice QCD}},
  \href{https://arxiv.org/abs/2002.12347}{{\ttfamily 2002.12347}}.

\bibitem{Crivellin:2020zul}
A.~Crivellin, M.~Hoferichter, C.~A. Manzari and M.~Montull, \emph{{Hadronic
  vacuum polarization: $(g-2)_\mu$ versus global electroweak fits}},
  \href{https://arxiv.org/abs/2003.04886}{{\ttfamily 2003.04886}}.

\bibitem{Keshavarzi:2020bfy}
A.~Keshavarzi, W.~J. Marciano, M.~Passera and A.~Sirlin, \emph{{The muon $g$-2
  and $\Delta \alpha$ connection}},
  \href{https://arxiv.org/abs/2006.12666}{{\ttfamily 2006.12666}}.

\bibitem{Aoyama:2020ynm}
T.~Aoyama et~al., \emph{{The anomalous magnetic moment of the muon in the
  Standard Model}},  \href{https://arxiv.org/abs/2006.04822}{{\ttfamily
  2006.04822}}.

\end{thebibliography}\endgroup
\end{document}